\documentstyle[psfig]{mn}

\newif\ifAMStwofonts  
  
\newcommand{\aap}{\rm A$\&$A}

\newcommand{\araa}{\rm ARAA}

\newcommand{\etal}{~et~al.~}                                
\newcommand{\y}{{\it y\ }}
\newcommand{\sen}{mJy/$\sqrt{\rm Hz}$}    
\newcommand{\Msolar}{M$_{\odot}$}                           
\newcommand{\Lsolar}{L$_{\odot}$}                           

\ifoldfss  
  \ifCUPmtlplainloaded \else  
    \NewTextAlphabet{textbfit} {cmbxti10} {}  
    \NewTextAlphabet{textbfss} {cmssbx10} {}  
    \NewMathAlphabet{mathbfit} {cmbxti10} {} 
    \NewMathAlphabet{mathbfss} {cmssbx10} {} 
  \fi  
  \ifAMStwofonts  
    \ifCUPmtlplainloaded \else  
      \NewSymbolFont{upmath} {eurm10}  
      \NewSymbolFont{AMSa} {msam10}  
      \NewMathSymbol{\upi}     {0}{upmath}{19}  
      \NewMathSymbol{\umu}     {0}{upmath}{16}  
      \NewMathSymbol{\upartial}{0}{upmath}{40}  
      \NewMathSymbol{\leqslant}{3}{AMSa}{36}  
      \NewMathSymbol{\geqslant}{3}{AMSa}{3E}

    \fi  
  \fi  
\fi 
  
\ifnfssone  
  \newmathalphabet{\mathit}  
  \addtoversion{normal}{\mathit}{cmr}{m}{it}  
  \addtoversion{bold}{\mathit}{cmr}{bx}{it}  
  \newmathalphabet{\mathbfit} 
  \addtoversion{normal}{\mathbfit}{cmr}{bx}{it}  
  \addtoversion{bold}{\mathbfit}{cmr}{bx}{it}  
  \newmathalphabet{\mathbfss} 
  \addtoversion{normal}{\mathbfss}{cmss}{bx}{n}  
  \addtoversion{bold}{\mathbfss}{cmss}{bx}{n}  
  \ifAMStwofonts  
    \ifCUPmtlplainloaded \else  
      \UseAMStwoboldmath  
      \makeatletter  
      \new@mathgroup\upmath@group  
      \define@mathgroup\mv@normal\upmath@group{eur}{m}{n}  
      \define@mathgroup\mv@bold\upmath@group{eur}{b}{n}  
      \edef\UPM{\hexnumber\upmath@group}  
      \new@mathgroup\amsa@group  
      \define@mathgroup\mv@normal\amsa@group{msa}{m}{n}  
      \define@mathgroup\mv@bold\amsa@group{msa}{m}{n}  
      \edef\AMSa{\hexnumber\amsa@group}  
      \makeatother  
      \mathchardef\upi="0\UPM19  
      \mathchardef\umu="0\UPM16  
      \mathchardef\upartial="0\UPM40  
      \mathchardef\leqslant="3\AMSa36  
      \mathchardef\geqslant="3\AMSa3E  
    \fi  
  \fi  
\fi 
  
\ifnfsstwo  
  \DeclareMathAlphabet{\mathbfit}{OT1}{cmr}{bx}{it}  
  \SetMathAlphabet\mathbfit{bold}{OT1}{cmr}{bx}{it}  
  \DeclareMathAlphabet{\mathbfss}{OT1}{cmss}{bx}{n}  
  \SetMathAlphabet\mathbfss{bold}{OT1}{cmss}{bx}{n}  
  \ifAMStwofonts  
    \ifCUPmtlplainloaded \else  
      \DeclareSymbolFont{UPM}{U}{eur}{m}{n}  
      \SetSymbolFont{UPM}{bold}{U}{eur}{b}{n}  
      \DeclareSymbolFont{AMSa}{U}{msa}{m}{n}  
      \DeclareMathSymbol{\upi}{0}{UPM}{"19}  
      \DeclareMathSymbol{\umu}{0}{UPM}{"16}  
      \DeclareMathSymbol{\upartial}{0}{UPM}{"40}  
      \DeclareMathSymbol{\leqslant}{3}{AMSa}{"36}  
      \DeclareMathSymbol{\geqslant}{3}{AMSa}{\"3E}  
    \fi  
  \fi  
\fi 
  
\ifCUPmtlplainloaded \else  
  \ifAMStwofonts \else 
    \def\upi{\pi}  
    \def\umu{\mu}  
    \def\upartial{\partial}  
  \fi  
\fi

\title[On the Detectability of the SZ Effect of Massive Young Galaxies]
{On the Detectability of the SZ Effect of Massive Young Galaxies}  
  
\author[D. Rosa--Gonz\'alez et al.]  
{Daniel Rosa--Gonz\'alez$^{1,2}$,
 Roberto Terlevich$^{1,3}$,   
Elena Terlevich$^1$\thanks{Visiting Fellow at IoA, UK},  Amancio Fria\c ca$^4$ \\  \\ 
{\LARGE and Enrique Gazta\~naga$^5$.}\\
$^1$ INAOE, Luis Enrique Erro 1. Tonantzintla, Puebla 72840. M\'exico.\\ 
$^2$ Astrophysics Group, Blackett Laboratory, Imperial College, Prince Consort Road, London SW7 2BW.\\  
$^3$ Institute of Astronomy, Madingley Road, CB3 OHA Cambridge, U.K.\\  
$^4$ Instituto Astron\^omico e Geof\'{\i}sico, USP, R. do Matao 1226, Cidade Universitaria, 05508-900, 
S\~ao Paulo, SP, Brazil.\\   
$^5$ Institut d'Estudis Espacials de Catalunya. Edifici Nexus, Gran
 Capita, 2-4, desp. 201, 08034 Barcelona, Spain}  

\date{Accepted  .  
      Received ;  
      in original form \today\  (ElGaSZ-v105)}  
  
\pagerange{\pageref{firstpage}--\pageref{lastpage}}  
\pubyear{2003}  
  
\begin{document}  
  
\maketitle  
  
\label{firstpage}  
  
\begin{abstract}

The Sunyaev-Zel'dovich (SZ) effect expected to be associated with 
massive star-formation activity 
produced during the formation of the most luminous bulges of normal galaxies is discussed.

Using 1-D chemohydrodynamical models for spheroidal galaxy evolution
we  show that during the early epochs of galaxy evolution 
the gas in massive events of star formation
may reach temperatures
and densities high enough to produce values of the comptonisation 
parameter \y comparable to those present in galaxy clusters.

In this scenario, we discuss the possibility of detection of the SZ signature in high
redshift starforming galaxies with 
the next generation of mm telescopes capable of arcsecond resolution 
and equipped with high sensitivity detectors.

We show how millimeter colour--colour diagrams or diagnostic diagrams could be used 
to distinguish between the dust emission and the SZ effect and suggest
the use of simultaneous  multifrequency observations to improve the chances of
detecting the SZ effect.

\end{abstract}  
  
\section{Introduction}

The SZ effect (Sunyaev \& Zel'dovich 1972) is produced when the cosmic 
microwave background (CMB) photons 
from the Rayleigh-Jeans region, due to the interaction with
the fast electrons, move to the Wien tail of the CMB spectrum thus
producing a unique spectral signature. This signature shows an increase 
with respect to the mean CMB brightness of
the observed intensity for wavelengths shorter than 1.34 mm and 
a decrease for wavelengths longer than 1.34 mm. 
This is known as the thermal SZ effect. 
Another component is the kinetic SZ effect, due 
to the interaction with CMB photons of plasma coherently moving with respect 
to the reference frame where the CMB is isotropic (e.g. Rephaeli 1995, Birkinshaw~1999, 
Church, Jaffe and Knox~2001).

The SZ effect constitutes a powerful tool to 
study the physical properties of the intercluster gas,
and when combined with  X-ray  observations,
also to  estimate  cosmological constants such as the 
Hubble constant H$_0$, the current mass density of the universe
$\Omega_{\rm M}$ or the cosmological constant $\Lambda$
(e.g. Majumdar 2001, Diego et al. 2002, Levine, Schulz \& White 2002,
Reese et al. 2002, Majumdar \& Mohr 2003).

It is understood that  in clusters the thermal SZ effect 
due to the interaction of CMB photons
with electrons with temperatures around 10$^8$K dominates over the kinetic SZ effect. 
This fact is evidenced by combining observations at 
1 mm where the thermal effect is close to the maximum, 
at 1.4 mm where it is almost zero 
and the kinetic SZ effect can be marginally detected
and at 2 mm where the thermal SZ has its minimum and is negative 
(e.g. Lamarre et al. 1998, Mauskopf et al. 2000).

The advent of a new generation of mm telescopes (LMT/GTM or ALMA, described in 
section~\ref{DetTel}) that combine relatively high angular resolution with 
high sensitivity will allow to explore the possibility of detecting 
the SZ effect in volumes much smaller than that of a cluster of galaxies.
Would it be possible for these new instruments to detect the SZ signature
in individual galaxies?

Natarajan \& Sigurdsson (1999), Rosa-Gonz\'alez et al.~(2000, 2001),
Majumdar, Nath and Chiba (2001) and Rosa-Gonz\'alez (2002) have indeed
explored the detactability of the SZ effect in individual 
galaxies.

Natarajan \& Sigurdsson (1999) studied the SZ effect due to 
galactic outflows powered by the mechanical energy provided by the 
accretion of matter to the central supermassive black hole in QSOs. 
Majumdar et al. (2001) propose the evolution of  supernova driven galactic winds
during the early stages of evolution of a normal galaxy,
as the cause of the distortion of the CMB radiation due to the kinetic SZ
effect. A similar model was originally proposed by Tegmark, Silk and Evrard 
(1993) to explain the absence of absorption lines in the observed spectra 
of high redshift quasars (Gunn-Peterson effect, Gunn \& Peterson~1965). 
The explosion model assumes an initial energy
input which  is equal to the 2\% of the total luminosity generated
in the supernova explosions during about 5$\times 10^7$ years. The
range of galaxy masses covered by this model goes from 
5$\times 10^7$\Msolar\ to  $10^{11}$\Msolar. 

Rosa-Gonz\'alez et al.~(2000, 2001) and Rosa-Gonz\'alez (2002) discussed the 
possibility of observing the  SZ effect in the centre  of young giant spheroidal galaxies. 
They include in the calculations both the thermal and the kinetic SZ effect. 

In  the {\it accepted} galaxy formation scenario described
by the hierarchical model (e.g. Navarro, Frenk \& White 1995),  
massive galaxies form after merging with smaller system that have
been formed in a previous epoch. 
However, there is observational evidence
that a significant number of L$^\ast$ galaxies  exist at redshift of 
about 5 and beyond (Hu, Cowie \& McMahon 1998, 
Stanway, Bunker \& McMahon 2003). 
Jimenez et al. (1999) discussed the possibility of the existence of 
massive elliptical galaxies at high redshift, contrary to 
early predictions of the hierarchical models of galaxy formation.
The study of massive galaxies at high redshift 
that could be detected through the unique SZ signature
could be used to constrain the  different galaxy 
formation scenarios. 

The SZ effect due to young galaxies 
together with radio sources (Holder 2002)
and dusty galaxies (Blain, Ivison \& Smail~1998, 
Blain~1998) are potential sources of noise in 
sensitive studies of SZ clusters and upcoming measurements of the CMB 
fluctuations (e.g. {\it FIRST}, {\it Plank}). Therefore, 
their characteristic emission and physical scale must be quantified. 

In this paper we explore in detail the possibility of detection of the SZ effect
in young galaxies undergoing a massive starformation event. We discuss the
confusion due to dust emission and suggest a diagnostic to segregate
between dust emission and SZ effect distortions.

\subsection{Thermal SZ effect}
\label{SZbasic}

The SZ effect describes the brightness change of the CMB due to the
inverse Compton scattering by hot electrons~\cite{1972Sunyaev}.

The change in intensity between the unaffected CMB radiation  and 
the radiation which go through the hot gas can be expressed by the integral,
\begin{equation}\label{eq:Int} 
\Delta I = \tau_e \frac{2h}{c^2}\int_{-\infty}^{+\infty} ds P(s)
 \left[ \frac{\nu_0^3}{\exp{\left(\frac{h\nu_0}{kT_{rad}}\right)}}  -  
        \frac{\nu^3}{\exp{\left(\frac{h\nu}{kT_{rad}}\right) }}
\right]
\end{equation}
where the energy of the CMB photons is given by $h\nu_o$, the energy of the 
photons which come out from the cluster is $h\nu$, 
$s$ is defined by $s=log(\nu/\nu_0)$, $\tau_e$ is 
the optical depth for scattering and P(s) is the probability that an
incident photon with energy $h\nu_o$ becomes a photon with an energy 
$h\nu$ \cite{1999Birkinshaw}. 

The integral in the Equation~\ref{eq:Int} can be calculated 
by using the non relativistic expression given by the Kompaneets approximation
\cite{1957Kompaneets},
\begin{equation}
\label{eq_SZII}
\Delta I(x) = h(x) y I_0
\end{equation}
\noindent
where the distortion of the photon spectrum, $h(x)$
is defined by
\begin{equation}\label{eq_SZ}
h(x) = x^4 \frac{e^x}{(e^x-1)^2} \left( x\ 
\hbox{coth}\,\frac{x}{2}-4\right)
\end{equation}
$x$ is the dimensionless frequency, $x=h\nu/(k T_{rad})$ with $T_{rad}=2.728$
the temperature of the CMB,
\begin{equation}\label{I0}
I_0 = \frac{2h}{c^2}\left(\frac{k_B T_{rad}}{h} \right)^3
\end{equation}
and the Comptonization parameter
``\y '' is  defined by the integral of the electron pressure 
 along the line of sight
\begin{equation}
\label{y_eq}
y=\sigma_T\int dl\ n_e\frac{k_B T_e}{m_e c^2}
\end{equation}
In this equation $\sigma_T$ is the Thomson scattering cross section, 
n$_e$ and $T_e$ are  the electron number density and the electron
temperature respectively, $k_B$ is the Boltzmann constant and $m_e c^2$
is the rest energy of an electron; 
(see Rybicki \& Lightman~1980 for a more detailed derivation of the different 
expressions described above and Birkinshaw 1999 for a recent review). 
The effect of the hot gas on the 
CMB spectrum is to produce a fractional change in the brightness 
of the CMB radiation. We remark that the amplitude of the 
SZ effect is  redshift-independent and depends only on the    
integrated  properties of all the hot gas along the line of sight.  
The SZ effect can be considered as a  
tool to detect high pressure electron gas over cosmological distances.

\begin{figure}
\setlength{\unitlength}{1cm}           
\begin{picture}(7,6)         
\put(-1.2,-6.7){\includegraphics{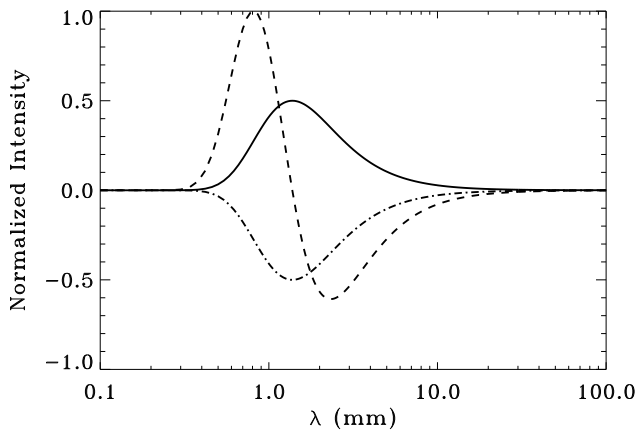}}
\end{picture}
\caption{\label{TerKin} The kinetic SZ effect (solid and dot-dashed lines) 
signature is compared to the thermal SZ effect (dashed line) 
for the case where the kinetic effect is  half of the thermal one.
The two cases of the kinetic SZ effect correspond to 
gas moving away from (dot-dashed line) and towards (solid line) the observer.}
\end{figure}

\subsection{Kinetic SZ effect}

\begin{figure*}
\setlength{\unitlength}{1cm}           
\begin{picture}(7,6)       
\put(-3.5,-11.5){\includegraphics{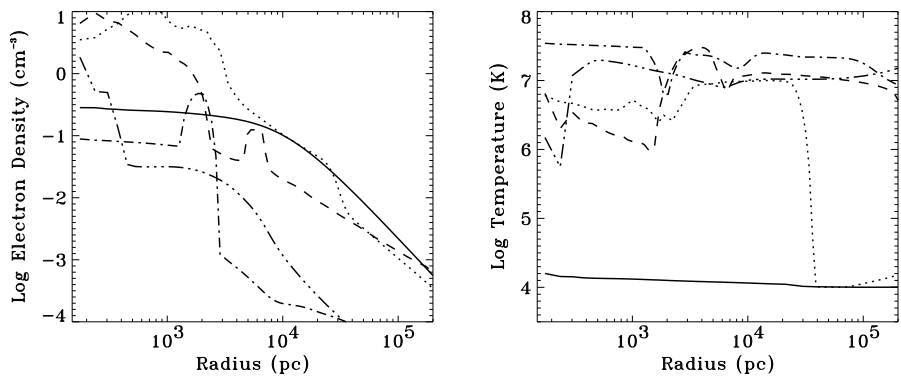}}
\end{picture}
\caption{\label{Profi} Electron density and temperature
profiles for the model of baryonic mass 2x10$^{12}$\Msolar. The different
lines correspond to different times: 20 Myr for the solid line,
600 Myr for the dotted line, 1 Gyr for the dashed line, 4 Gyr for the
dot--dashed 
line and 11 Gyr for the dot-dot-dashed line.}
\end{figure*}

The kinetic SZ effect, as already mentioned, is due to the movement of the 
plasma with respect to the CMB rest frame. 
For a shell of matter which is moving with a radial velocity $v$ 
with respect to the Hubble flow, the change of the intensity 
due to the kinetic SZ effect ($\Delta I_{kin}$) is given by,
\begin{equation}
\label{KinEq}
\Delta I_{kin} = - \frac{v}{c} \tau_e I_0 \frac{x^4e^x}{(e^x - 1)^2}
\end{equation}
where $\tau_e$ is the scattering optical depth,  $I_0$  is given 
by equation~\ref{I0} and $x$ is the dimensionless frequency.

At low frequencies, the ratio between the kinetic SZ 
effect ($\Delta I_{kin}$) to the thermal SZ effect ($\Delta I_{th}$) 
is independent of the scattering optical depth and can be calculated by, 
\begin{equation}
\label{KTratio}
\frac{\Delta I_{kin}}{\Delta I_{th}} = \frac{1}{2}\frac{v}{c}\frac{m_e c^2}{kT_e}
\end{equation}
where $v$ is the velocity of the flow, $m_e c^2$ is the electron rest energy
and $T_e$ is the gas temperature~\cite{1999Birkinshaw}.
The signatures of  the thermal and kinetic SZ effects
are plotted in Figure~\ref{TerKin}.

\section{The galaxy formation model}

In order to estimate the amplitude and time evolution of the SZ effect
for individual galaxies, it is necessary to know the time and radial dependence
of the gas temperature, density and velocity.
We have restricted our study to those stages of galaxy evolution
characterized by a high injection rate of mass and energy into the galaxy ISM
(which happens to occur during the first few Gyr of the galaxy evolution). 
We have further assumed that this occurs during
a maximum in the star formation rate produced by any of the known mechanisms.

The evolution and radial profiles mentioned above
(Figure \ref{Profi})
have been obtained from a chemohydrodynamical model for evolution 
of elliptical galaxies (Fria\c ca \& Terlevich 1998, hereafter FT98).
The model combines
multi-zone chemical evolution with 1-D hydrodynamics to
calculate the evolution of chemical abundances in gas and stars,
and also the temperature, density and velocity profiles of the gas.
In this way, we can calculate
the \y profiles which are used to estimate the flux collected
by a given beam size at a given frequency.

In the FT98 model, a single massive dark halo
hosts baryonic gas that falls toward the centre of the dark halo
and will subsequently form stars.
The dark halo is given by a static mass density distribution
$\rho_h(r)=\rho_{h0}[1+ (r/r_h)^2]^{-1}$,
where $\rho_{h0}$ is the halo central density and $r_h$ is the halo core 
radius.
The gas and the stars exchange mass through star formation
and stellar mass losses (supernovae, planetary nebulae, and stellar winds).
Both the stellar distribution and the dark halo are truncated 
at a common tidal radius $r_t$.
The system, assumed to be spherical, is 
subdivided in several spherical zones and the 
hydrodynamical evolution of its ISM is 
calculated. The equations of chemical evolution for each 
zone are then solved taking into account the gas flow,
and the evolution of the chemical abundances is obtained.
A total of $\approx$100 star generations are stored during 13 Gyr 
for chemical evolution calculations. 
We assume that at a given radius $r$ and time $t$,
the specific star formation rate $\tilde \nu(r,t)$
follows a power-law function of gas density ($\rho$):
$\tilde \nu(r,t)=\nu (\rho/\rho_0)^{1/2}$,
where $\rho_0$ is the initial average gas density inside $r_h$,
and $\nu$ is the normalization of the star formation law.
We include inhibition of star formation for expanding gas ($\nabla.u>0$)
or when the density is too low, and, therefore, the cooling is inefficient
(i.e., for a cooling time $t_{coo}=(3/2)k_B T/\mu m_H \Lambda(T) \rho$
longer than the dynamical time $t_{dyn}=(3\pi/16\,G\,\rho)^{1/2}$)
by multiplying $\tilde \nu$ as defined above by the inhibition factors
$(1+t_{dyn}\,{\rm max}(0,\nabla.u))^{-1}$ and $(1+t_{coo}/t_{dyn})^{-1}$.
A characteristic of these models is that several episodes of inflow
and outflow occur simultaneously at different radii.
The chemodynamical model for spheroids was used to investigate 
the relation between young elliptical galaxies and QSO activity (FT98),
the absence of passively evolving elliptical galaxies in deep surveys
(Jimenez et al. 1999),
Lyman Break Galaxies (Fria\c ca $\&$ Terlevich 1999),
Blue Core Spheroids (Fria\c ca $\&$ Terlevich 2001),
the coupled spheroid and black hole formation (Archibald et al. 2002),
and the link between DLAs and dwarf galaxies (Lanfranchi \& Fria\c ca 2003).

The models are parameterized
according to the (initial) baryonic mass inside the tidal radius,
$M_G=M_g+M_*$, $r_h$, $r_t$,
and the ratio of the halo to the (initial) luminous mass, $M_h/M_G$.
We have investigated a grid of runs with $M_G$ between $10^{11}$ 
and $5\times10^{12}$ \Msolar,
and $r_h$ in the range $2.5-15$ kpc (see FT98).
We set $r_t=28r_h$ and $M_h/M_G=5$. 
That value is compatible with the recent WMAP results where 
the baryon density $\Omega_{\rm b}=~0.047~\pm~0.006$ and 
the matter density $\Omega_{\rm M}=~0.29~\pm~0.07$ (Spergel et al. 2003).
Smaller (larger) values  of the ratio dark matter  versus baryonic mass 
do not significantly alter the observed fluxes,
but due to the shallower (deeper) potential well, the galactic winds 
will appear earlier (later).
For example, in the model with $M_G=10^{12}$ \Msolar,
the galactic wind occurs at 1.51, 1.55 and 1.67 Gyr,
for $M_h/M_G=3$, 5 and 7.5 respectively.  

The chemical evolution is driven by the stellar winds, planetary
nebulae and SN phase which produce the enrichment of the ISM. 
New  generation of stars  form in the more metal-rich medium. 
We do not assume instantaneous recycling approximation for the chemical
enrichment, but instead we take into account
the delays for gas restoring from the stars
due to the main-sequence lifetimes.
Instantaneous mixing with the ISM is assumed for the stellar ejecta.
We use metallicity dependent yields 
for SNe II, SNe Ia, and intermediate mass stars (IMS):
the SNe II yields of Woosley $\&$ Weaver (1995), 
for metallicities Z/Z$_{\odot}$ = 0, 10$^{-4}$, 10$^{-2}$, 10$^{-1}$ and 1;
SNIa yields from Iwamoto et al. (1999) --- their models W7  and W70 
(progenitor metallicity Z=Z$_{\odot}$ and Z=0, respectively);
the yields for IMS ($0.8 - 8$ \Msolar),
with initial Z=0.001, 0.004, 0.008, 0.02 and 0.4,
from van den Hoek \& Groenewegen (1997) (their variable $\eta_{AGB}$ case).
For more details of the nucleosynthesis prescriptions, see FT98.
The models start with an entirely  gaseous protogalaxy with
primordial chemical abundances ($Y=0.24$, $Z=0$).

In the FT98 models there is  self-consistency of
the hydrodynamics, chemical evolution and atomic physics,
since the cooling function is evaluated based on the actual
chemical abundances obtained from the chemo-dynamical modelling.
For the sake of simplicity,
our cooling function is a function only of the abundances
of O and Fe, which are the main coolants for $T>10^5$ K.
In the calculation of the cooling function, 
the abundances of elements other
than Fe and O have been scaled to the O abundance as
$n_i=n_{i,P}+(n_{i,\odot}-n_{i,P}) n_O/n_{O,\odot}$,
where $n_i$ is the ISM abundance by number of the element $i$, 
$n_{i,P}$ its primordial abundance and $n_{i,\odot}$ its solar abundance
[the photospheric values of Holweger (2001) for N and O , and
the meteoritic values of Grevesse \& Sauval (1998) for the other elements].

It is important to note that the FT98 models 
describe  the evolution of isolated galaxies and do not take into account
effects such as mergers that play an important role in
structure formation within  hierarchical galaxy formation models.
However, the FT98 models can be considered as a description of
the evolution of the hot gas after a major star forming event produced by 
interaction or merger.
In addition,
while the present chemodynamical model accounts for complex flow patterns, 
with inflow in some regions and outflow in other regions,
the model does not take into account asymmetric fluxes
due to the assumed spherical symmetry.
Despite the limitations due to the spherical symmetry, 
the detailed predictions of this model are useful 
due to the good radial resolution of the galaxy and the realistic  treatment
of the star formation and proper calculation of the chemical evolution
of the gas and stars. 
The FT98 model represents a complementary tool
to the semi-analytic models of galaxy formation,
which currently adopt very simplified recipes of star-formation
but with coarser space resolution
(e.g. Sommerville \& Primack 1999, Benson et al. 2001).

An important aspect relevant to the present calculation is that
the models provide an estimate of the size of the hot gas region. 
This is central for deriving the expected flux because 
the effective \y parameter (which is the average of the 
\y parameters within a given beam size, Section~\ref{SZinGG}) 
depends strongly on the size of the telescope beam.

\subsection{The SZ effect in  massive young  galaxies.}\label{SZinGG}

The FT98 models predict that the hot gas in massive star forming 
regions with ages between 10$^8$ and  10$^9$ years  can reach 
temperatures exceeding 10$^7$K and densities higher than 1 cm$^{-3}$
within a radius of a few kpc.
These temperatures and densities imply values of the Comptonization
parameter  \y larger than 10$^{-4}$.
The  maximum temperatures of the gas in the models ($\sim$ 1 keV) 
are well below the relativistic limit ($\sim$ 500 keV),
and, therefore, we use the
non-relativistic approximation (e.g.~Birkinshaw~1999) to
estimate the SZ effect (Equation~\ref{eq_SZII}).
Notice that the central \y parameter 
is calculated for a region of a few hundred parsecs across. 
Only in this region is the \y parameter comparable 
to those observed in galaxy clusters.

\begin{figure}
\setlength{\unitlength}{1cm}           
\begin{picture}(7,6)         
\put(-1.2,-6.7){\includegraphics{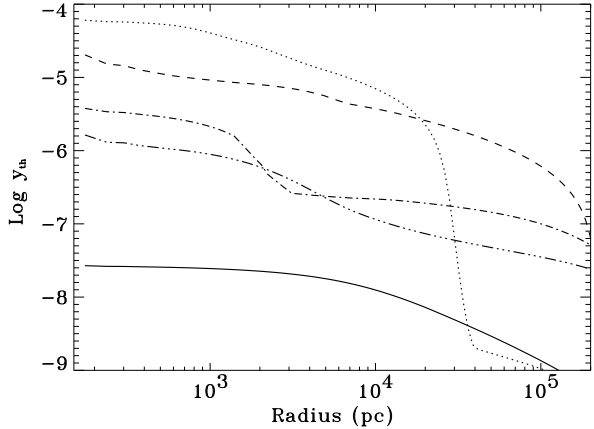}}
\end{picture}
\caption{\label{RatioTK} 
The thermal \y parameter as a function of radius. 
Different lines correspond to different times: 20 Myr for the solid line,
600 Myr for the dotted line, 1 Gyr for the dashed line, 4 Gyr for the
dot--dashed 
line and 11 Gyr for the dot-dot-dashed line. The baryonic mass 
of the galaxy is 2$\times 10^{12}$\Msolar.}
\end{figure}

Figure~\ref{RatioTK} shows the variation of the thermal parameter
($y_{th}$) with radius.
In the beginning of the calculations 
(the $t=20$ Myr lines in Figures 2 and 3),
the gas has not yet settled in the potential well of the galaxy
and its density is low and temperature is cold everywhere
(the model initial conditions assume gas at $T=10^4$ K),
and, as a consequence, \y is very small.
As the model evolves, the gas falls towards the centre and is compressed,
giving rise to shocks that 
rapidly heat the gas in the core to approximately 
the virial temperature of the system ($T\sim 10^7$ K). 
Then, a highly efficient star formation is occurring throughout the galaxy
and a young stellar population is rapidly built up.
Following the initial violent star formation burst, 
the first Type II SNe appear and heat the ISM.
The star formation is vigorous and intense during the first
$\sim 0.5\;$Gyr, when nearly half of the stellar population is formed.  
In Figures 2 and 3, we see that at $t=0.6$ Gyr, there is enough
gas at temperatures around $T=10^7$ K 
to maintain values of \y close to $10^{-4}$ in the central kpc.
At 1 Gyr since the start of the calculations, the star formation
gas consumed most of the gas and the SNII (and SNIa) heating 
is past its maximum.
Although the gas temperature is higher than before, 
its density has dropped along most of the galaxy,
resulting in a significant decrease of \y.
As the galaxy evolves, the star formation rate decreases,
the main heating source of the ISM is provided by Type Ia supernovae.
Eventually, the thermal energy of the gas is enough to overcome
the escape velocity, and the remaining gas is expelled from the galaxy
by galactic winds.  
At this point, the primary burst of star formation ceases.
The curves at $t=4$ and 11 Gyr illustrate the post-galactic wind stage,
when the gas temperature is high, but the low density implies a low {\it y}.

The \y parameter is smaller in the external layers where the pressure
is lower, therefore the resulting \y parameter averaged within the telescope 
beam is expected to be lower than the central one. This effect can be 
quantified by 
an effective $\y$ parameter ($y_{eff}$) defined as the average of
the \y parameter  in the different shells weighted by the
corresponding projected area and convolved with the telescope beam 
profile,

\begin{equation}
y_{eff} = \left(\frac{1}{\theta^2} \int y \, d\theta\right)*B(\theta)
\end{equation}

\noindent
where $\theta$ is the size of the telescope beam and 
$B(\theta)$ is the  telescope beam  profile.

\begin{figure}
\setlength{\unitlength}{1cm}           
\begin{picture}(7,6)         
\put(-1.2,-6.7){\includegraphics{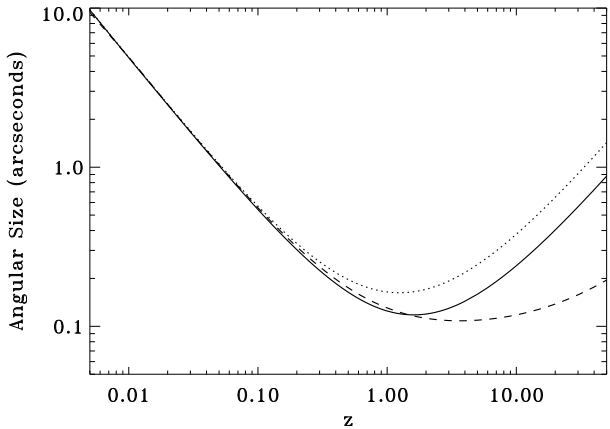}}
\end{picture}
\caption{\label{1kpc} Angular size corresponding to a physical 
size of 1 kpc for different cosmologies. The solid line is the result for a 
universe with $\Omega_{\rm M}=0.3$ and $\Omega_\Lambda=0.7$. 
The dotted line is for the case of $\Omega_{\rm M}=1$ and
$\Omega_\Lambda=0$ and
the dashed line is for    $\Omega_{\rm M}=0.05$ and  $\Omega_\Lambda=0$.}
\end{figure}

In reality, the telescope beam will cover a 
variable fraction of the galaxy depending on its intrinsic
size, its redshift  and the adopted cosmology (Figure~\ref{1kpc}).
In order to translate physical sizes to angular sizes, 
we adopted a cosmological model given by $\Omega_{\rm M}=0.3$, $\Omega_\Lambda=0.7$ and 
h=0.7.

\begin{figure*}
\setlength{\unitlength}{1cm}           
\begin{picture}(7,6)         
\put(-1.5,-6.5){\includegraphics{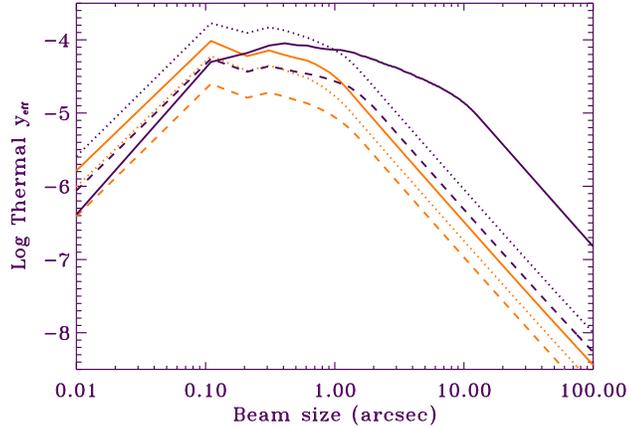}}
\end{picture}
\caption{\label{yTheta}  
The lines represent
the variation of the beam averaged \y  parameter with
beam size for a fixed time of $4\times 10^8$ years,
 corresponding to galaxies 
of baryonic masses of 10$^{11}$\Msolar\ (grey dashed line), 
2$\times10^{11}$\Msolar\ (grey dotted line),
5$\times10^{11}$\Msolar\ (grey solid line), 
10$^{12}$\Msolar\ (black dashed line), 
2$\times10^{12}$\Msolar\ (black dotted line) and
5$\times10^{12}$M$_\odot$\ (black solid line).
}
\end{figure*}

To illustrate the effect of beam dilution we show in
Figure ~\ref{yTheta} the change of $y_{eff}$  with respect 
to the beam angle for models at 400 Myr. The results are similar for 
different times. The behaviour of $y_{eff}$  can be 
analysed  in three separate cases, according to the beam size:
\begin{itemize}
\item  {\bf Small beam}, $y_{eff}$ increases as $\theta^2$, where $\theta$ is
the size of the beam. In this case the beam size is smaller than the first 
radius of the model at the given time, so the value of \y increases until
the telescope beam reaches the inner radius of the model galaxy.
\item {\bf Medium beam}, for  beam angles similar to the inner radius,  
the model obtains the maximum value of \y which corresponds to the central
one ({\it y}$_c$). The integrated
pressure in the external layer is lower than in the inner ones, hence
increasing the beam size produces a slight decrease of $y_{eff}$. 
\item  {\bf Large beam}, when the size of the beam 
is larger than the region where the
values of \y are significant  the averaged \y goes down as  $\theta^{-2}$.
Notice that for the most massive galaxy (baryonic mass  5x10$^{12}$\Msolar) 
the fast decline is reached after  an angle of about 10\arcsec\ but for the 
low mass galaxies the 
fast decline starts when the angle is about 1\arcsec. 
\end{itemize}

\begin{figure*}
\setlength{\unitlength}{1cm}           
\begin{picture}(7,6)         
\put(-6,-6.5){\includegraphics{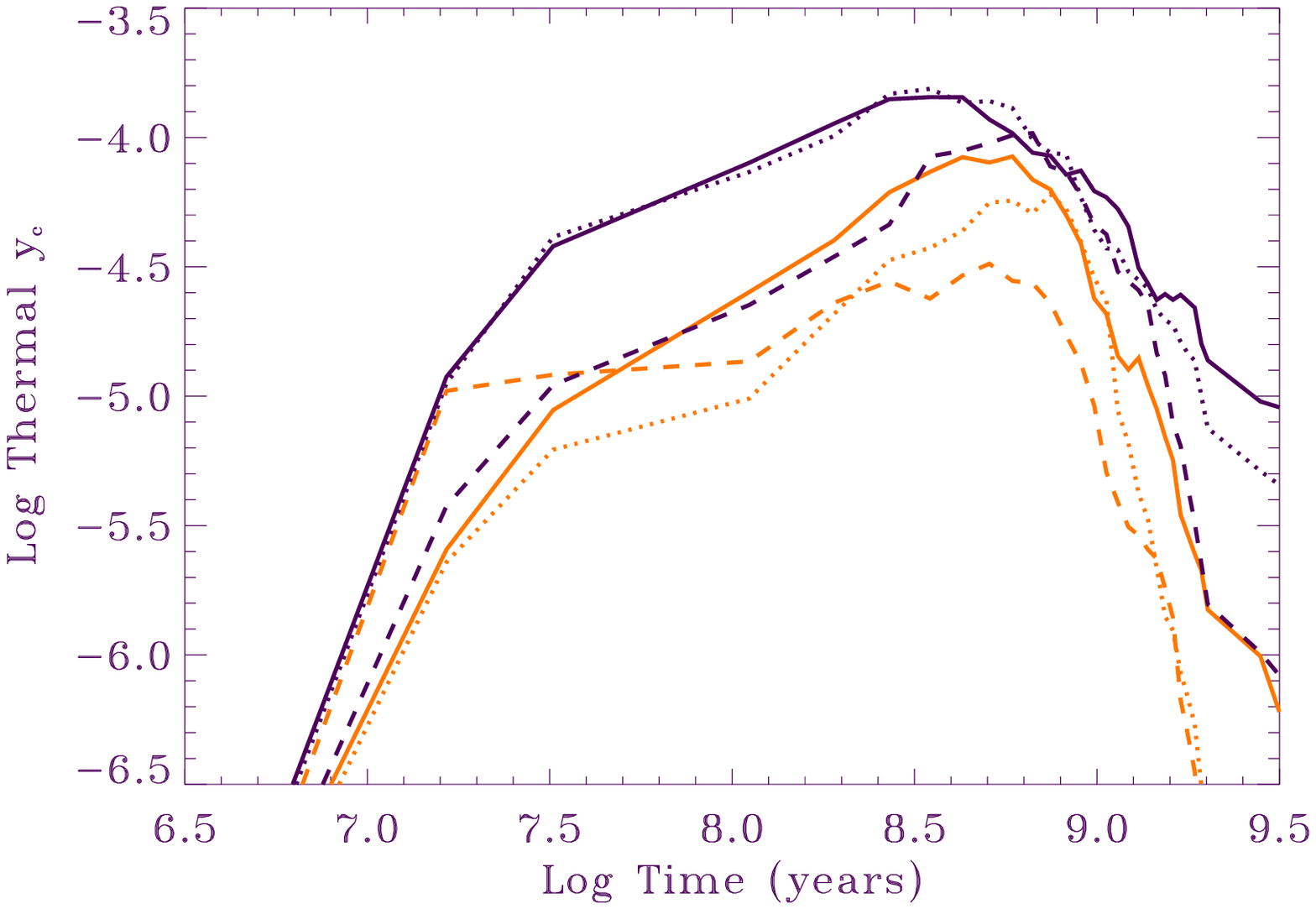}}
\put(2.5,-6.5){\includegraphics{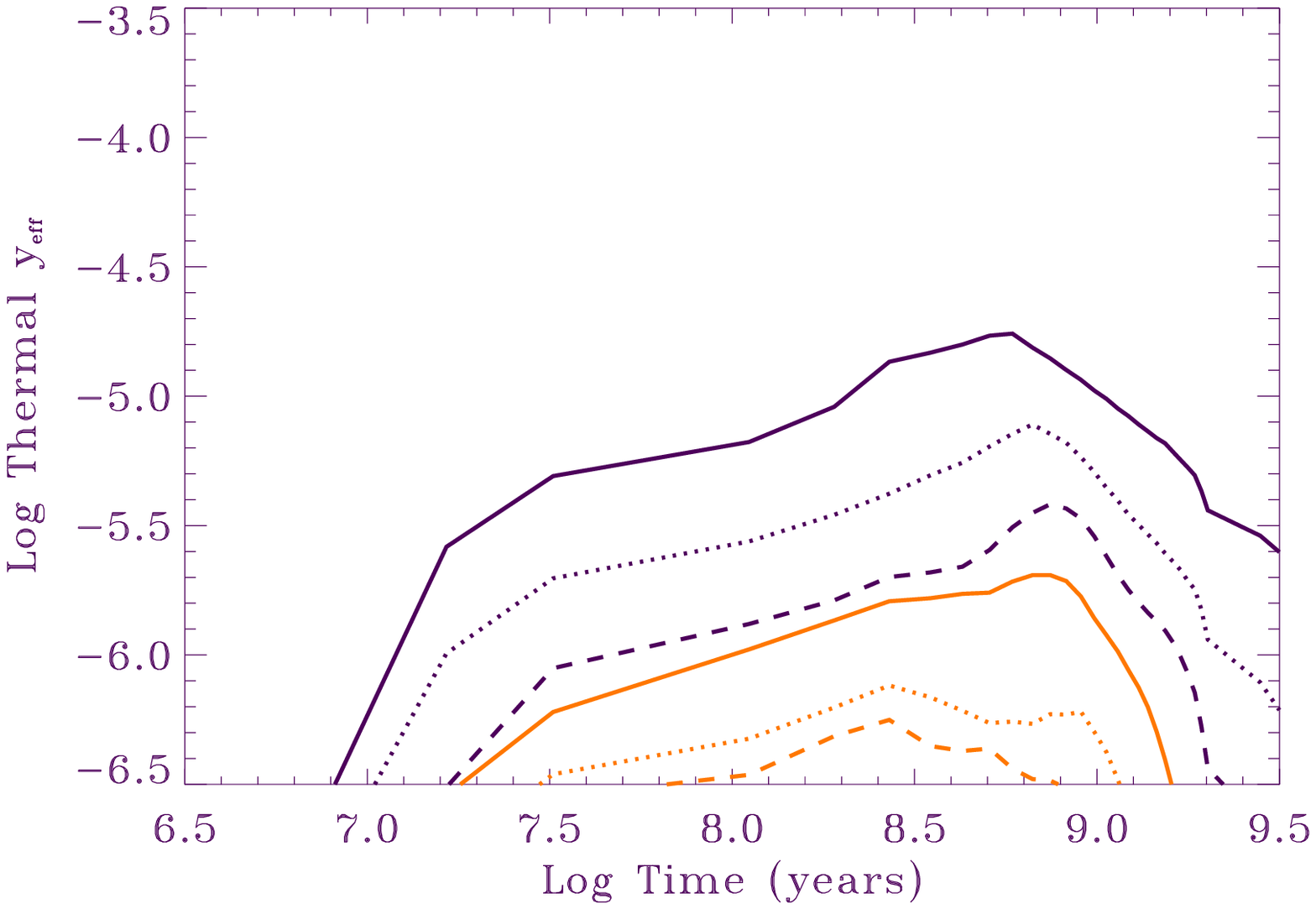}}
\end{picture}
\caption{\label{Yes} Evolution of the
\y parameter. The left panel is the evolution of the 
central thermal \y parameter ($y_c$).
The right
panel is the effective thermal  \y parameter ($y_{eff}$) 
assuming a beam size of 5\arcsec.
The different lines
correspond to different galaxy masses, coded as in Figure~\ref{yTheta}.
}
\end{figure*}

Figure~\ref{Yes} shows that for the massive 
galaxies,  the comptonisation parameter
reaches values higher than  3$\times$10$^{-5}$.  
In the case of  quasar outflows,  Natarajan \& Sigurdsson (1999)
estimated a value for the comptonisation parameter
of about 5$\times$10$^{-5}$. 
Similar values of the comptonisation parameter 
were estimated for the case of early galactic winds (Majumdar et al. 2001). 

However, the dilution effect  produces a significant decrease 
of the  \y parameter and consequently a smaller
expected flux (Figure~\ref{Yes}).  
Using the output of the models, we obtained the
effective \y parameter as a function of time and beam angle.
The behaviour of the central \y and the 
effective \y parameter for the case of a beam 
of 5\arcsec\ is plotted in Figure~\ref{Yes}. The central \y
parameters  
have the maxima  between 150$\times10^6$ years and 10$^9$ years, 
independently of the baryonic mass of the galaxy. The maximum values go
from 2$\times10^{-4}$ for galaxies with baryonic  masses 
of 5$\times 10^{12}$ and 2$\times 10^{12}$\Msolar\ down to 
3$\times 10^{-5}$ for galaxies with baryonic masses of 10$^{11}$\Msolar.
The effective \y within an angle of 5\arcsec is about 10 times lower
than the central one (Figure~\ref{Yes}). 
The given times in Figures~\ref{Yes},~\ref{TotFlux} and~\ref{R12} correspond to 
galaxies which started to collapse at redshift 50. 
Within this initial conditions,  the maximum SZ amplitude is reached 
at redshifts between 12 and 5. 

\subsection{The kinetic SZ effect in young galaxies}
\label{SZK}

There are two processes
that could produce a significant kinetic SZ effect in young galaxies. One is  the
infall of the primordial gas to the center of the galaxy and the other
is the outflow produced by  SN explosions and winds from massive young stars. 

The calculations of the thermal SZ effect presented in the previous section,
are for a spherical galaxy 
(the models by FT98 on which the calculations
are based, assume  spherical symmetry). For the case of the 
kinetic SZ effect of a spherical galaxy with isotropic winds, the total 
contribution to the kinetic \y parameter becomes zero. In reality, 
the geometry of galactic winds tend to be more bi-conical than spherical.
If there is asymmetry between the two cones, 
there will be a net kinetic SZ effect.
Even within the context of our spherically symmetric calculations,
we can estimate the kinetic SZ effect, in the case of extreme asymmetry,
by considering the contribution of only one hemisphere of the galaxy.

\begin{figure}
\setlength{\unitlength}{1cm}           
\begin{picture}(7,6)         
\put(-1.25,-6.5){\includegraphics{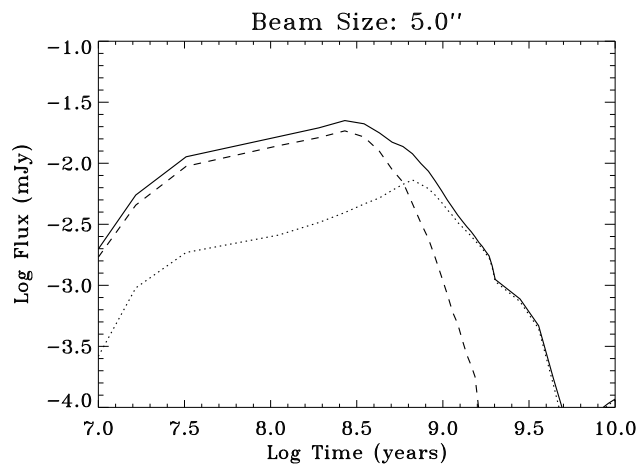}}
\end{picture}
\caption{\label{TotFlux} The  total flux (solid line)
due to the sum of the thermal (dotted line) and
kinetic (dashed line) SZ  effects for a galaxy of 2$\times 10^{12}$\Msolar\ (baryonic mass).
We assumed that the kinetic effect has a
positive sign so the corresponding flux is added directly to the flux due
to the thermal SZ.
The beam  angle (5\arcsec) corresponds to  observations
at 1 mm with a  50 meter telescope.}
\end{figure}

In this gross approximation,
the kinetic effect is dominant at early phases of the galaxy evolution
(less than 0.5 Gyr) due to the infall of gas to the centre of the galaxy.
During this epoch the gas is relatively cold with temperatures
of about 10$^4$K favouring the kinetic effect over the thermal one. 
During this early phase the SZ kinetic effect could be 
several times higher than the thermal effect (Figure \ref{TotFlux}).
In the central region of the galaxy the thermal effect dominates 
due to the high temperatures and densities and as a result the 
central $y_{th}$ parameter is higher than the kinetic component.

Notice, however, that an assumption of extremely asymmetric infall
is probably less justified than that of asymmetric wind. 
However, when the galactic wind is operating ($t>1$ Gyr)
due to its low density,
the implied kinetic  SZ effect would be much less than the
thermal SZ effect. Moreover at these late times, the galaxy
is far from the time  of maximum SZ effect
($t \sim 0.5$ Gyr).

\section{Telescopes and detectors}
\label{DetTel}

There are a few future  mm and sub-mm facilities
that may be capable of detecting the SZ effect produced by individual 
galaxies.

One is the 
Atacama  Large Millimeter Array (ALMA), a project lead by institutions 
of Europe, Japan and North America. ALMA will be an array of 
64 antennas of 12 m in diameter getting baselines up to 10 km. It is
expected that ALMA will cover the wavelength range from 
8 mm to 350~$\mu$m~\cite{1999Blain}. It will be situated in the 
Atacama desert in Chile.

Another one is the Large Millimeter Telescope (LMT/GTM), a binational project 
between UMASS (USA) and INAOE (M\'exico),  with  a 
single dish antenna of 50 m diameter. It will be situated on the Cerro
la Negra in M\'exico.
In the case of observations of the continuum, that is the case of 
the SZ effect, the LMT/GTM will be equipped with BOLOCAM, 
a sensitive wide field camera that will operate between 3 and 1 mm.
At 1 mm on the LMT/GTM, BOLOCAM will have a projected pixel size 
of 5\arcsec, three times smaller than SCUBA on the JCMT
observing at 850~$\mu$m~\cite{1998Holland}.

We have also included the characteristics of the Green Bank
Telescope (GBT) a 100 m single dish which is planned to be able to work at 
3 mm in the next few years~\cite{2002Dicker}.

\begin{table*}
\begin{center}\caption{\label{sensi} Main characteristics of the most sensitive
telescopes at the mm range. 
The upper value of 10 km for ALMA correspond to the maximum separation
between two antennas. 
The mapping speed ($\ast$) is measured in
$\frac{{\rm arcminute}^2}{\rm{hour\ mJy}^2}$.}
\begin{tabular}{|c|c|c|c|c|c|r|}\hline
Telescope          & Camera  &  Wavelength &Angular  Resolution& Sensitivity   & FOV          & Mapping Speed \\
and diameter       & \       &   (mm)      &(arcsec) &  (\sen)       & (arcmin$^2$) &($\ast$)\ \ \ \ \ \ \ \ \ \\ \hline\hline
JCMT (15 m)        & SCUBA   &   0.85      & 15 &    90         & 2            & 0.9  \\
GBT (100 m)        & 3mm--Camera     &   3.00  & 8   &    0.2          & 0.3  &  27000 \\ 
LMT/GTM (50 m)     & BOLOCAM &   1.2      &  6  &  2           & 2           &   1800 \\
ALMA ($<$ 10 km)   &   --    &   1.30      & 0.03  &  0.46        & 0.16          & -- \\ \hline
\end{tabular}
\end{center}
\end{table*}

The mapping speed is defined as the ratio between the field of view (FOV)
and the  square of the sensitivity. High mapping speed is 
necessary to complete surveys covering 
big portions of the sky and to detect the fainter sources.

Table~\ref{sensi}
shows the sensitivity, FOV and mapping speed 
for the  different telescopes.
The sensitivity of BOLOCAM in the LMT/GTM is estimated from Glenn~\etal 1998. 
The sensitivity and FOV of ALMA are from Blain (1999).
Notice that ALMA is not designed to perform wide area surveys 
(so it does not have an associated mapping speed).

We are 
interested in the detection of the SZ effect in young massive 
galaxies  that at high  redshift may have a few arcseconds in size 
and an estimated flux  of several $\mu$Jy (See Section~\ref{SZinGG}).  
Table~\ref{sensi} shows that  only with the next generation of mm telescopes
it may be possible to detect them.

\section{The effect of  dust emission}

Dust emission is a dominant component in the observed multifrequency spectrum
of galaxies over a wide range of redshifts.
Given the small signal associated with the SZ of individual galaxies, dust emission
will probably dominate the mm spectrum making difficult the detection of the
SZ signature. At high redshift the peak of the dust emission ($\sim$100~$\mu$m) moves
to sub-mm and mm wavelengths (e.g. Rowan-Robinson et al. 1997, Hughes et al. 1998) 
i.e. over the region where the SZ effect has its maximum.

We estimated the relative amplitude of the flux due to the SZ effect and that 
due to the dust by comparing  the mm flux due to dust emission in M82, 
a well known starforming galaxy, 
with the expected flux due to the SZ effect.

M82 is a nearby ($\sim 3.2$ Mpc) starburst galaxy with an  infrared
luminosity of L$(8-1000 \mu m)=3\times 10^{10}$\Lsolar\ which represents the
84\% of the total galaxy luminosity~\cite{1988Telesco}.
 Hughes, Gear and Robson (1994) showed that 
the spectral distribution of energy  in the sub-mm range of M82 is
well fitted by a grey body law with a temperature of 48 K and an
emissivity index of 1.3.

The dust emission is,

\begin{equation}
\label{M2F}
F_\nu = M_{dust} k_d^{rest}B(\nu^{rest},T) \frac{\displaystyle 
1 + z}{\displaystyle D_L^2} 
\end{equation}
\noindent 
where $z$ is the redshift of the source, 
$k_d^{rest}$ is the rest-frequency mass absorption coefficient (e.g. Rowan-Robinson 1986),  
$B(\nu^{rest},T)$ is the rest-frequency
value of the Planck function for dust at temperature $T$, 
$D_L$ is the luminosity distance and  ($M_{dust}$)
is the total mass of dust
(Hughes, Dunlop and Rawlings 1997).
For the rest-frequency mass absorption coefficient ($k_d^{rest}$)
we adopted the value of 0.15 m$^2$kg$^{-1}$ as suggested by
Hughes et al. (1997).
The dependence of $k_d$ with wavelength is given by $k_d\propto
\lambda^{- \beta}$, where $\beta$ is the emissivity index of the dust
grains.

\begin{figure}
\setlength{\unitlength}{1cm}           
\begin{picture}(7,6.)         
\put(-1.25,-6.5){\includegraphics{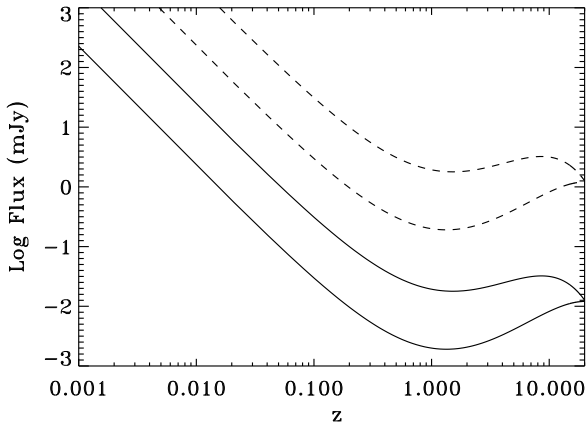}}
\end{picture}
\caption{\label{M82} The solid lines are the expected flux for 
M82 at 2 mm (lower solid line) and at 1 mm (upper solid line)
assuming a dust temperature of 50 K and $\beta=$1.5. The dashed
lines are the expected flux for a galaxy 100 times more luminous than 
M82. The lower dashed line corresponds to fluxes at 2 mm and 
the upper dashed line, to fluxes at 1 mm.}
\end{figure}

Observations of galaxies with different sources 
of heating indicate an average  dust grain 
temperature of about 50 K
(e.g. Chini \etal 1989, Chini, Kreysa and Biermann 1989,  
Hughes \etal  1993, Hughes et al. 1994). 
In what follows we assume a dust temperature of 50 K and an
emissivity index of 1.5 given by Hughes et al.~(1997).
The dust mass of M82 is about  2~x~10$^6$\Msolar~\cite{1994Hughes}.
We also compare the flux due to dust in galaxies
hundred times brighter than M82 (Arp~220-like luminosity, Figure~\ref{M82})
to the flux due to the SZ effect (Figure~\ref{TotFlux}).
We must keep in mind that Arp~220 is an extreme case of an 
ultraluminous infrared galaxy (ULIRG) with an infrared luminosity of 
about $\sim 10^{12}$\Lsolar which  corresponds to  98\% of the total 
luminosity. Combining the observed flux at 850$\mu$m
(F(850$\mu$m)=792$\pm$26 mJy)
given by Lisenfeld, Isaak and Hills (2000) with Equation~\ref{M2F}  we 
estimate for Arp~220 a dust mass of about 10$^8$\Msolar. A value of 
2$\times$10$^8$\Msolar for the dust mass is given by 
Dunne et al. (2000) assuming a dust temperature of 42.2 K and an
emissivity index of 1.2.

Galaxies as M82 have an expected flux at 1 mm
of about 20 $\mu$Jy for redshifts between 1 and 10 and about 5 $\mu$Jy 
for observations at 2 mm (100 times higher for the 
case of an Arp~220-like galaxy). Notice that these values are 
similar to the maximum values of the fluxes due to SZ emission 
for massive galaxies with baryonic masses higher   
than 10$^{11}$\Msolar (Figure~\ref{TotFlux}). 
The estimate of total fluxes due to dust, does not take into account the 
relative size of the galaxy since the SZ flux  plotted in
Figure~\ref{TotFlux}
corresponds only to the flux estimated inside a beam size of 5\arcsec.
At high redshift ($z>$3) the whole galaxy  will fill the 5\arcsec\ beam, 
but probably will be  resolved by interferometric observations.
The other fact is that the dust content of M82 does not change with
time,  therefore the fluxes  in Figure~\ref{M82}  can be considered
upper limits for galaxies with lower metal content or less evolved than M82.

\section{Differential mapping and the mm Colour-colour diagram}

The particular spectral signature of the SZ effect, i.e.
a positive maximum at about 800 $\mu$m and a negative minimum  
at around 2 mm plus its non-dependence on
redshift can be used to maximize its detection against the
dust emission from the same source.

\begin{figure}
\setlength{\unitlength}{1cm}           
\begin{picture}(7,6)         
\put(-1.25,-6.5){\includegraphics{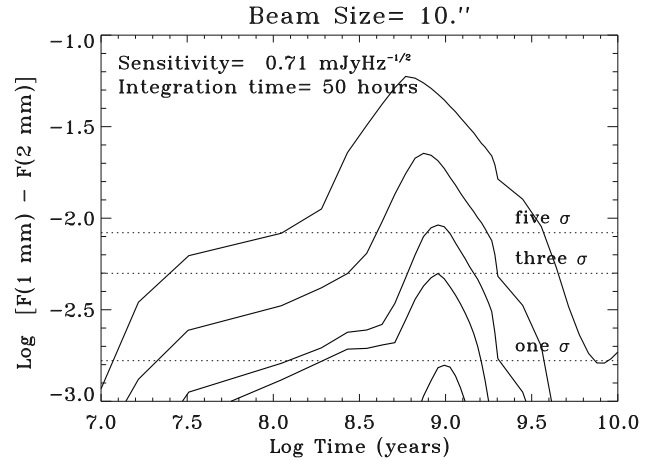}}
\end{picture}
\caption{\label{R12} Difference between the fluxes at 1 mm and
at 2 mm. The different lines correspond 
to different galaxy masses;  
from the bottom to the top the baryonic masses are 2$\times$10$^{11}$
5$\times$10$^{11}$, 10$^{12}$, 2$\times$10$^{12}$ and
5$\times$ 10$^{12}$\Msolar.
Only the thermal SZ was included in this figure. The horizontal dotted lines 
are the estimated rms assuming the  sensitivity and integration time shown 
in the upper left corner.}
\end{figure}

Using the fact that at 2 mm the SZ has a negative value,  
the difference between 
the flux at 1 mm and the flux at 2 mm  (Figure~\ref{R12})  is almost
two  times  the signal given by the individual values (Figure~\ref{TotFlux}).
Notice that subtracting the signal at 1 mm from the signal at 2 mm 
will also cause a decrease in the flux due to sources dominated 
by dust emission therefore producing a map biased to SZ sources.
Source confusion due to dust emission should also be  reduced.
This differential mapping method will also help in removing the sky signal as 
at these low resolutions the atmosphere 
emission at 1 mm and at 2 mm are  correlated. In a single 
observation with a wide area camera,  the sky signal is removed 
by using the fact that some of the pixels are viewing the blank
sky. Subtracting the time average level of these blank pixels 
can increase the signal to noise by a factor of 3. This is possible
because the source structure is constant in time while  the 
sky vary over the array on time scales of several seconds~\cite{1998Holland}.
The use of simultaneous measurements at different wavelengths allows to
reduce the sky noise due to a better sampling of the fast
sky fluctuations. Also it is possible to  
separate different components of the image (i.e. detector noise, sky,
astrophysical signal) without assuming any atmospheric
spectrum by solving a set of coupling equations and obtaining a model of the 
sky brightness distribution~\cite{2003Phil}.

To distinguish between  galaxies which are dominated
by the dust emission and galaxies which are dominated 
by the SZ effect, we have explored the use of mm colour-colour diagnostic diagrams.

\begin{figure}
\setlength{\unitlength}{1cm}           
\begin{picture}(7,6)         
\put(-.5,-.50){\includegraphics{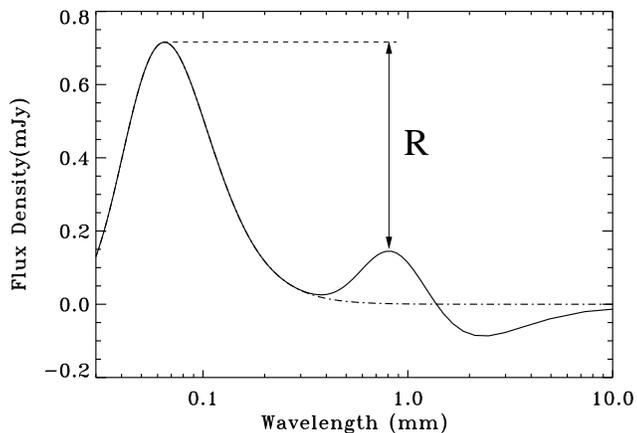}}
\end{picture}
\caption{\label{dlplanckm}
The dash--dotted line shows the spectrum of a grey body ($\beta=2$) 
at 50K.
The  grey body  emission has been matched to 5 times the
maximum of the SZ emission.
The solid line shows the SZ
effect with $y=10^{-4}$ added over the grey dust.
We assumed a beam  of 5\arcsec. {\bf R} is the ratio between 
the maximum of the  dust emission and  the maximum  of the SZ effect as 
defined in the text.}
\end{figure}

Let us consider the emission of an object  composed by 
the thermal SZ effect with 
a fixed comptonization parameter of 10$^{-4}$ plus a grey body. 
The amplitude of the
grey body emission is given by changing {\bf R}  which 
is the ratio between the maximum of the grey body emission
and the maximum of the SZ emission  (Figure~\ref{dlplanckm}).

\begin{figure*}
\setlength{\unitlength}{1cm}           
\begin{picture}(7,11)         
\put(-6.,-0.5){\includegraphics{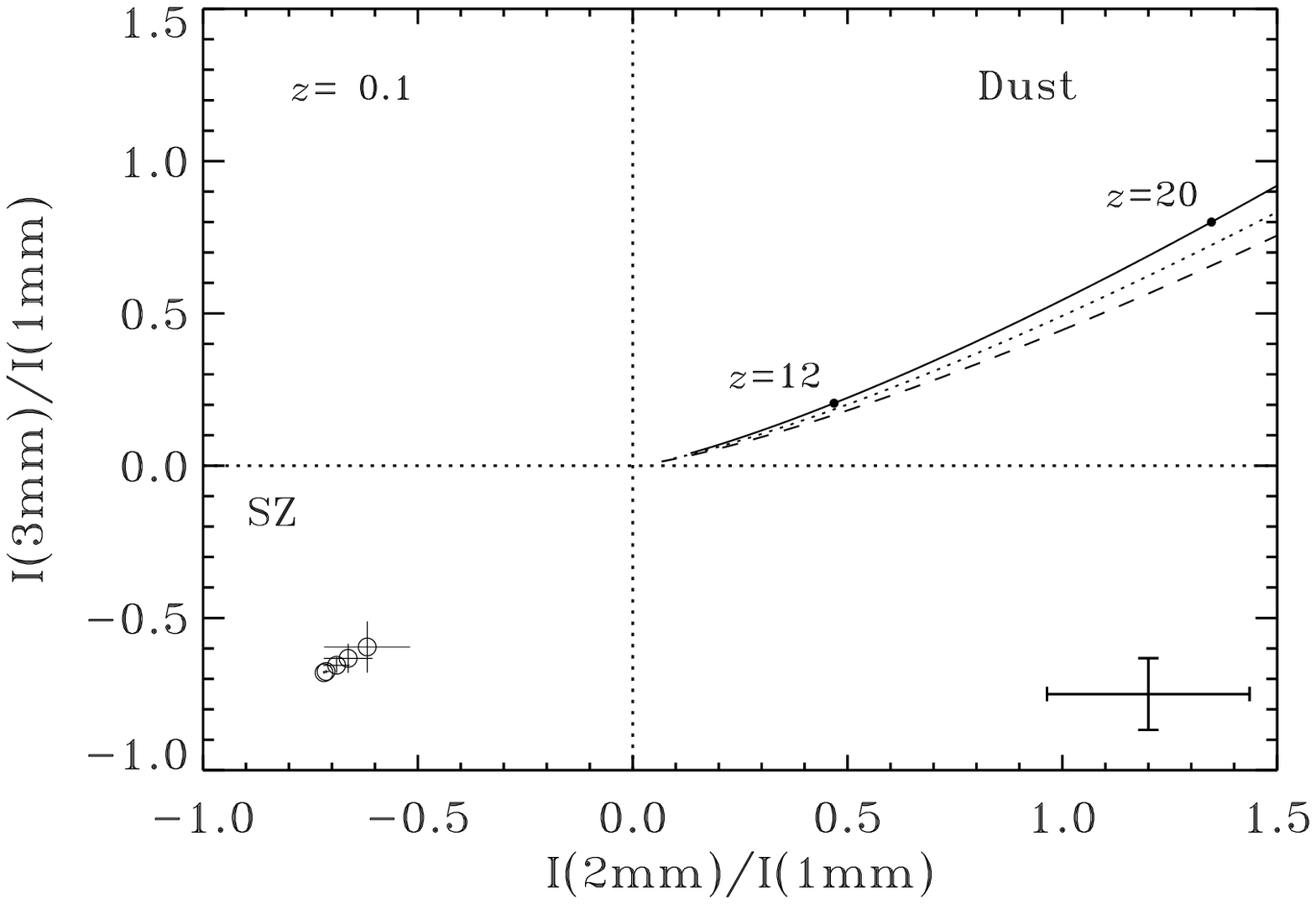}}
\put(3.,-0.5){\includegraphics{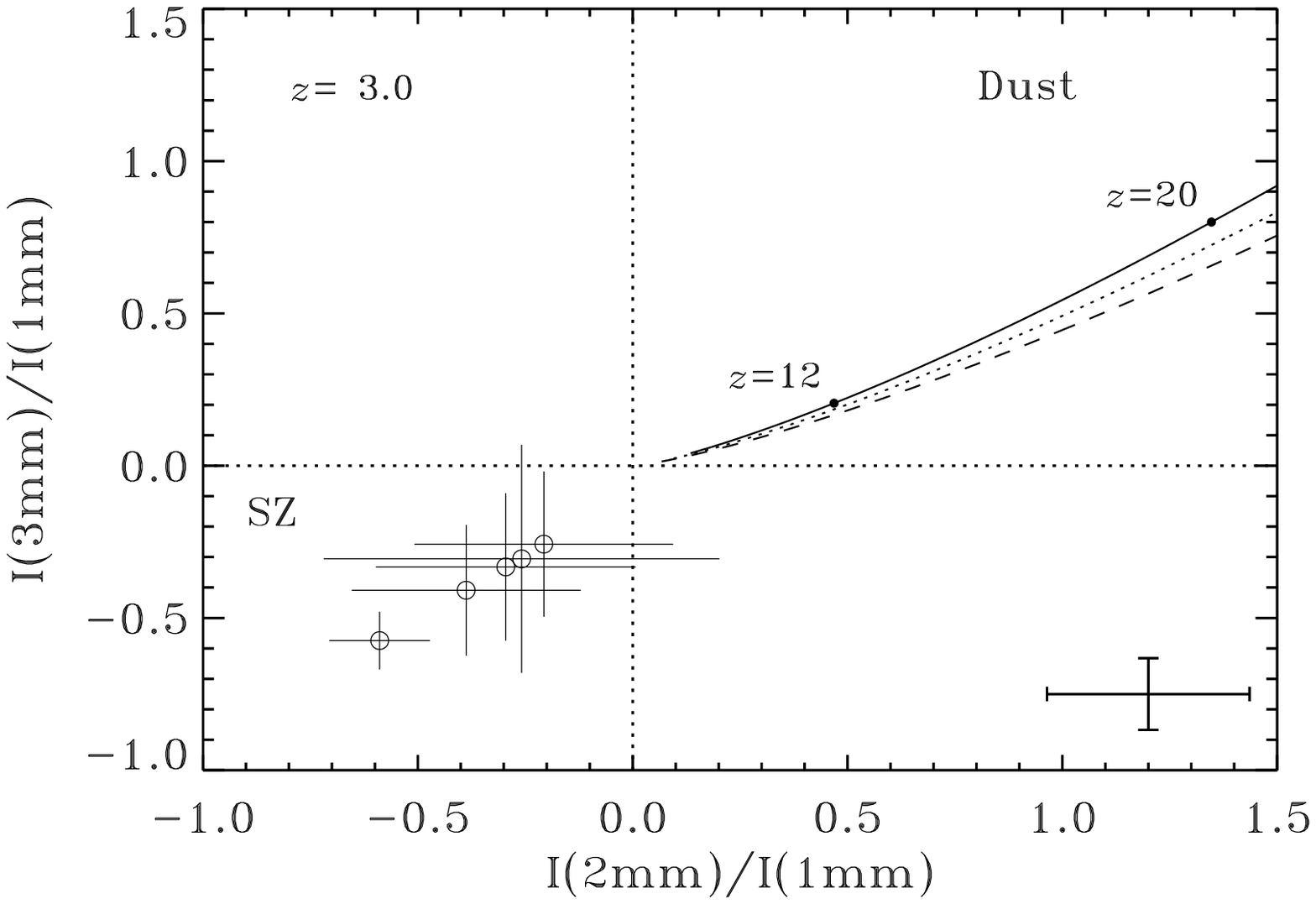}}
 \put(-6.,- 6.){\includegraphics{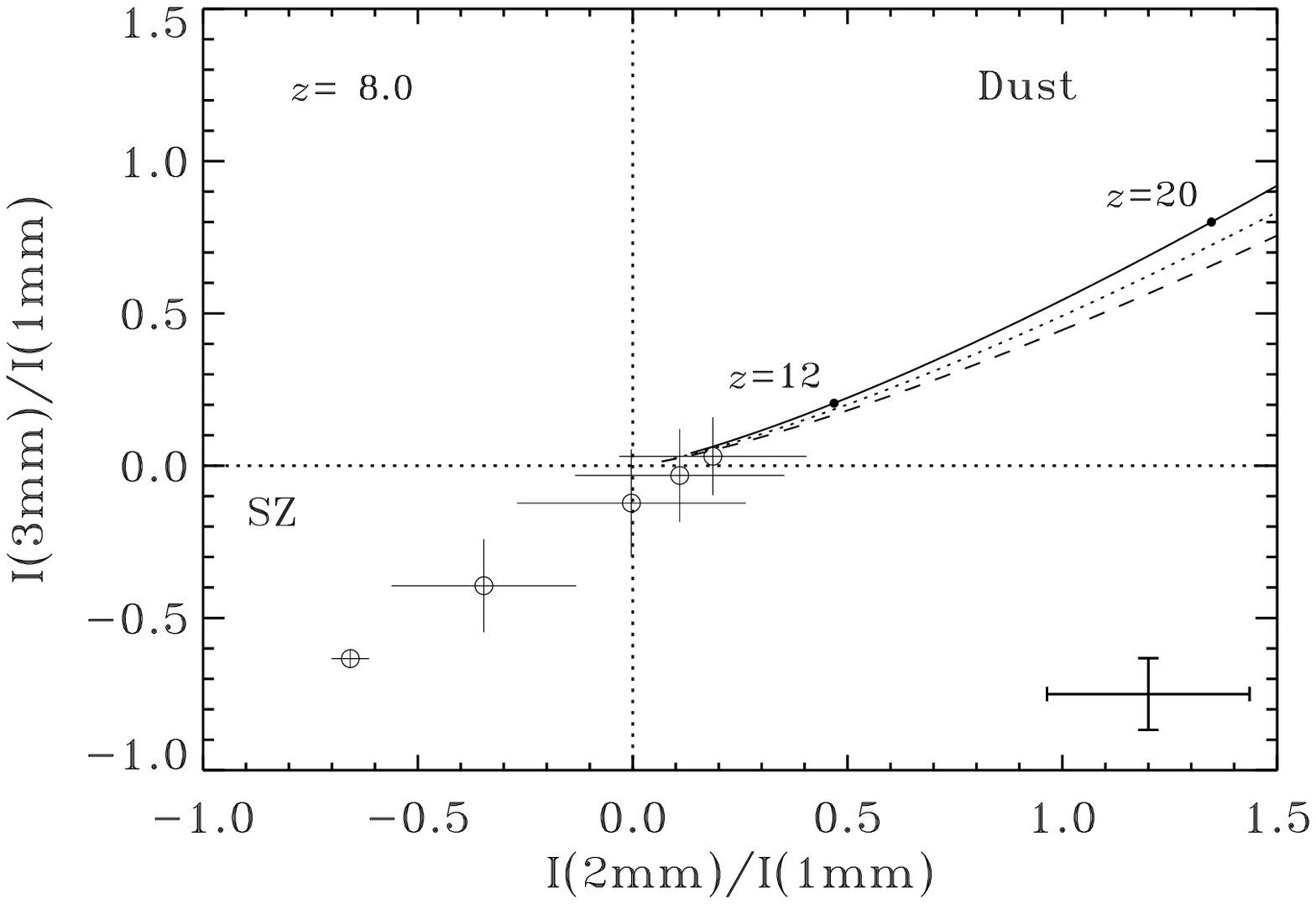}}
\put(3.,- 6.){\includegraphics{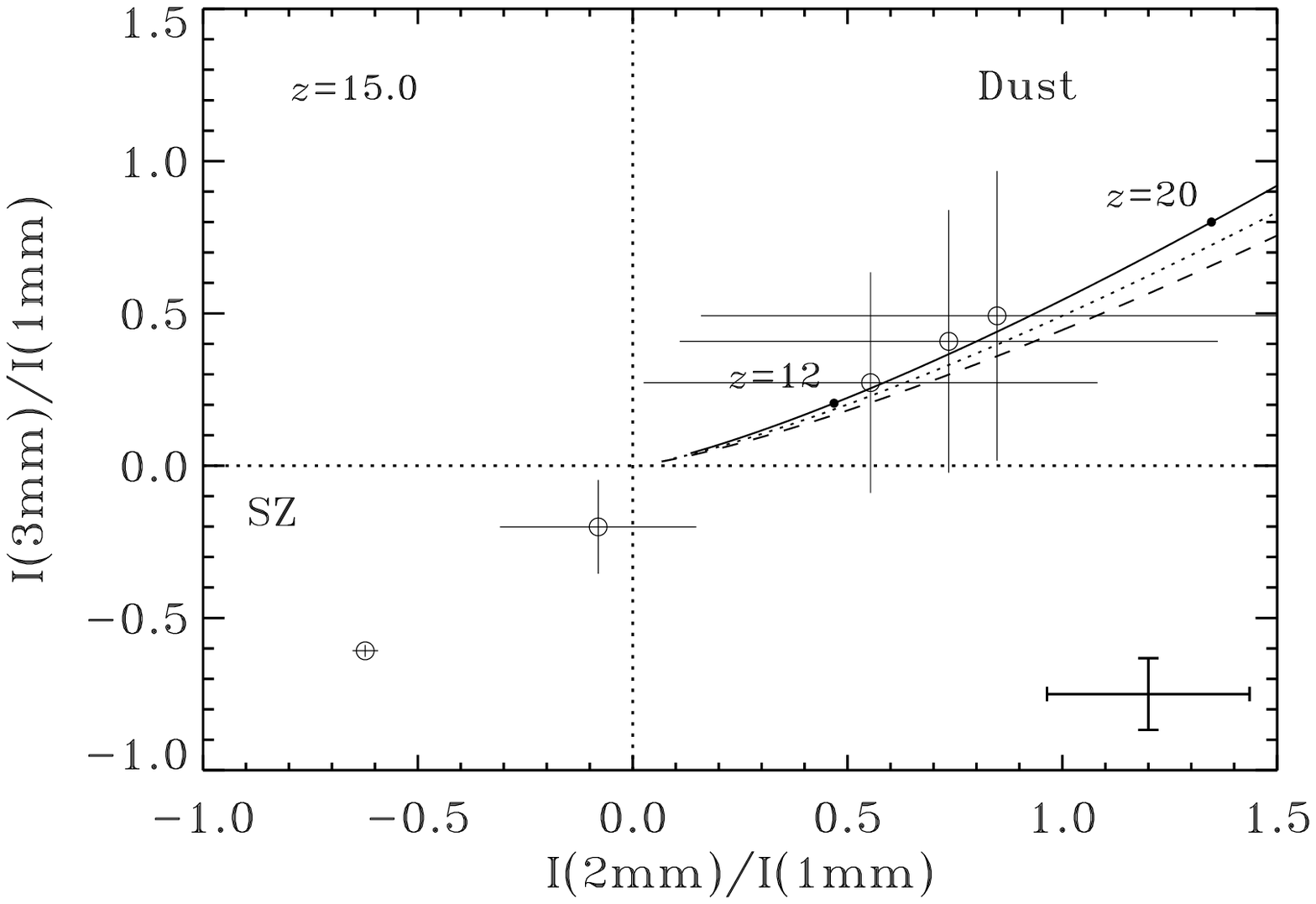}}
\end{picture}
\caption{\label{ColCol}
The different panels represent a set of 
mm colour--colour diagram for sources with both the emission of the SZ effect
and the emission from the dust.
The different panels correspond to galaxies at different redshifts:
$z$=0.1 (top left), $z$=3 (top right), $z$=8 (bottom left) and  $z$=15 (bottom right).
For each individual panel, the different points correspond to
the adopted values of the ratio $R$ =0.1,1.,5.,10.,20.
The amplitude of the bars are calculated by varying the
grey body temperature from 35 to 65K and the emissivity parameter from 1 to 2.
The lines correspond to T=50K grey body models with  emissivity parameters
1 (solid line), 1.5 (dotted line) and 2 (dashed line). The values corresponding to $z$=12 and $z$=20 
are indicated. The cross in the bottom right corner corresponds to  
observational errors for  3$\sigma$ detection  in all three bands.}
\end{figure*}

Figure~\ref{ColCol} represents the plane I(2 mm)/I(1 mm) 
vs.~I(3 mm)/I(1 mm) given by the ratios between the intensities
at the given wavelengths. The intensity of the SZ effect is fixed
by adopting a \y parameter of 10$^{-4}$.  The 
dust emission is given by a grey body with temperature of 50 K 
and an emissivity of $\beta$=1.5. 
The amplitude of the dust emission is calculated by varying {\bf R} 
({\bf R}=0.1,1.,5.,10.,20.).
The error bars were estimated by fixing the values of 
{\bf R}  and redshift but  allowing the temperature to vary between 
35 and 65K and the emissivity parameter between 1 and 2. 

Figure~\ref{ColCol} allows us to examine the behaviour of the mm colours
for different redshift ranges:
\begin{itemize} 
\item Low redshift galaxies ($z \la 1$). In this case
the peak of the grey body is far away from the mm range where the signal 
due to the SZ is dominant (between 1 mm and 3 mm). For {\bf R}=0.1 we obtain 
almost the (negative) pure SZ colour (I(2 mm)/I(1 mm)=--0.72, 
I(3 mm)/I(1 mm)=--0.68). For higher values of {\bf R} the 
Rayleigh-Jeans  tail of the grey body radiation starts to contaminate 
the SZ spectra producing a shift towards the right top of
the I(2 mm)/I(1 mm) vs. I(3 mm)/I(1 mm) plane,
but, for these redshifts, the displacement is modest.

\item Intermediate redshift galaxies ($2 \la z \la 5$). This case
is similar to the previous one but the grey emission on top of the SZ
is more important, producing a 
larger displacement of the points to the right top corner of the diagram,
while the mm colours remain negative.

\item High redshift galaxies ($z > 5$). In
this case the R-J part of the grey body emission has covered the 
SZ effect emission at 2 mm, producing in some cases  positive I(2 mm)/I(1 mm) and 
I(3 mm)/I(1 mm) ratios.
\end{itemize} 

Figure~\ref{ColCol} indicates that all points in the bottom left
part of the diagram have a dominant  SZ component. This diagnostic 
seems valid up to redshift of several. The accuracy or reliability of the 
diagnostic will clearly depend on the size of the error bars that in turn
depend on the amplitude of the signal and sensitivity of the telescope.
All in all it seems that the new facilities would permit to start the 
exploration of at least the largest star-forming systems at $z<12$ 
even if they show moderate dust emission.
Inside the estimated error bars, sources dominated by the 
SZ effect fall clearly in the distinct quarter of the diagramme where 
both
the 2 to 1 and 3 to 1~mm intensity ratios are negative.
This result is very robust in the sense 
that sources dominated by the SZ are going to 
be separated from the dust emission independently of redshift. The
confusion is higher for high redshift galaxies with  
the dust emission comparable to the SZ emission.


The diagnostic diagram presented in Figure~\ref{ColCol} would 
be partially covered by future deep surveys at mm wavelengths, 
showing high redshift sources dominated either by dust or by SZ emission.
Because of the expected delay between the burst event 
and the production of metals and dust, 
our diagnostic diagramme  can potentially be used to constrain the sources 
and time scales for dust and metal production in the early universe 
(Morgan \& Edmunds 2003).


To further advance in the study of the evolution of the 
SZ and dust emission in a self-consistent manner, 
to check in more detail the possibility of observing the SZ through 
individual galaxies
with the next generation of mm telescopes and to estimate the
contribution of SZ sources to the  background radiation,
we will present in a forthcoming paper a set of simulated  mm  maps  
including sources dominated by SZ and
those dominated by dust emission (with different clustering
properties) plus the comparison 
of our predictions on the SZ background with those made by 
Aghanim, Balland  \& Silk (2000) based on the SZ due to 
black holes-seeded proto-galaxies.

\section{Conclusions}

In this paper we have explored the possibility that the central 
comptonization \y parameter in young spheroidal 
galaxies or galaxies undergoing a massive starforming event may
reach values comparable to those  observed in galaxy clusters.

However due to the poor angular resolution-sensitivity combination  of present day mm facilities,
the detection of the 
SZ effect in individual galaxies may only be  possible with the next generation of
high angular resolution - high sensitivity mm telescopes (e.g. GTM/LMT or ALMA) and then 
probably only in the
most massive systems capable of producing a 800~$\mu$m flux of more than a few $\mu$Jy for
periods of time of several hundred megayears.

We have included instrumental noise in our calculations, but
neglected the noise contribution caused by high redshift
millimeter background point sources (e.g. Blain 1998,
Hughes et al. 1998).  The confusion due to unresolved sources 
depends on the actual details (i.e. resolution) of the observation,
the (unknown) emission radial profile of these sources and the
extrapolation of the number counts to fainter fluxes. 
Blain et al. (2000)  give an estimation of the confusion limit
of about  10~$\mu$Jy for the case of  observations at 
1 mm and a beam size of 5\arcsec. 
This value is comparable to the 
expected SZ flux of galaxies with baryonic masses of 
2$\times 10^{12}$\Msolar\ and an age of $\sim 3\times 10^{8}$ years (Figure~\ref{TotFlux}).
Confusion will be a major problem when combining independent observations at different
frequencies but its effect will be greatly reduced using  instruments that simultaneously 
observe at three selected frequencies.


A major difficulty in the detection of the SZ effect in an individual
galaxy will be its confusion with the dust emission originated in the same 
star formation event.
To solve the dust contamination problem we suggest to exploit the 
particular spectral signature of the  SZ spectrum and maximize
its segregation from dust emission by using  
simultaneous multifrequency observations or differential detection   
at 1, 2 and 3 mm.  
In the diagnostic diagram presented in Figure~\ref{ColCol}
there is a wide region centred at I(2 mm)/I(1 mm)=--0.3 and 
I(3 mm)/I(1 mm)=--0.3 where the thermal SZ emission
can be clearly separated from the dust emission over a wide range of 
redshifts.

Due to its particular spectrum the detection of the SZ signature in
individual galaxies and the discrimination 
between sources dominated by the SZ effect from those sources dominated by 
dust emission would be enhanced  by the use of differential mapping 
systems that simultaneously 
observe at three selected frequencies (Section~\ref{DetTel}). 
Additional advantages of such systems will be the lowering of the minimum detected 
flux due to a better sampling of the time correlated atmospheric fluctuations 
(i.e. reducing the sky noise) and the relative absence of source confusion
generated by dust emitting sources.

New generation of mm cameras such as SCUBA-2 \footnote{http:
$\backslash\backslash$www.roe.ac.uk$\backslash$atc$\backslash$projects$\backslash$scuba\_two}
or SPEED \footnote{http:
$\backslash\backslash$www.astro.umass.edu$\backslash\sim$fcrao$\backslash$instrumentation} -- which will 
have the capability of observing simultaneously in more than a band -- 
would be ideal to detect compact and faint SZ sources.



\section{Acknowledgments}  

Daniel Rosa Gonz\'alez gratefully acknowledges financial support
from  CONACYT, the Mexican Research Council, as part of ET research grant  
\#~32186-E. At present DRG is supported by POE, a European Research Training
Network. Discussions with David Hughes, Richard Ellis, Manolis Plionis and Guillermo
Tenorio-Tagle together with useful suggestions from an anonymous referee greatly improved this work. 


\bsp  
\label{lastpage}  
\end{document}